\documentclass[rnote]{aa}
\usepackage[varg]{txfonts}
\usepackage{graphicx}
\usepackage{natbib}
\usepackage{lscape}
\begin{document}
\title{Young Stellar Object candidates toward \\
       the Orion region selected from GALEX}
\author{Nestor Sanchez \and
        Ana In\'es G\'omez de Castro \and \\
        F\'atima Lopez-Martinez \and
        Javier L\'opez-Santiago}
\institute{S. D. Astronom\'ia y Geodesia,
           Fac. CC. Matem\'aticas,\\
           Universidad Complutense de Madrid,
           28040 Madrid, Spain}

\date{Submitted: 18 July 2014. Revised: 24 October 2014.}

\abstract{We analyze 359 ultraviolet tiles from the All Sky Imaging
Survey of the space mission GALEX covering roughly 400 square degrees
toward the Orion star-forming region. There is a total of 1,555,174
ultraviolet sources that were cross-matched with others catalogs
(2MASS, UCAC4, SDSS, DENIS, CMC15 and WISE) to produce a list
of 290,717 reliable sources with a wide range of photometric
information. Using different color selection
criteria we identify 111 Young Stellar Object
candidates showing both ultraviolet and infrared excesses, of
which 81 are new identifications. We discuss the spatial
distribution, the spectral energy distributions and other
physical properties of these stars. Their properties are,
in general, compatible with those expected for T~Tauri stars.
This population of TTS candidates is widely dispersed around
the Orion molecular cloud.}
\titlerunning{Young Stellar Objects toward Orion}
\authorrunning{S\'anchez et al.}
\maketitle

\section{Introduction}

The current paradigm for low-mass star formation categorizes
Young Stellar Objects (YSO) into four general classes: Class~0
sources are faint central protostars surrounded by a massive
envelope, Class~I sources are more evolved protostars with both
circumstellar disks and envelopes, Class~II sources (which include
the Classical T~Tauri stars) correspond to
pre-main sequence stars with significant
levels of circumstellar material in an accretion disk configuration,
and Class~III (or Weak-lined T~Tauri stars) refers to pre-main
sequence stars that have very low or even no accretion \citep[see,
for example, the reviews by][]{Shu87,Ber89,Mck07}. Deriving and
comparing the physical properties among these different evolutive
phases is necessary for a full understanding of disk evolution and
both planet and star formation.
T~Tauri stars (TTSs) are particularly interesting because of the great
variety of physical processes occurring in them. The central star and
its circumstellar disk interact via the magnetic field so the gas is
channeled through the magnetic field lines and accelerated forming
an accretion shock on the stellar surface. The magnetic interaction
between star and disk is complex, simultaneously producing accretion
flows, collimated outflows and winds \citep[see reviews
in][]{Bou07,Gom13}.

From the observational point of view, the distinction between
Classical and Weak-lined TTSs is based on the presence and
strength of H$\alpha$ emission and Lithium absorption lines
\citep{Bas91,Mar97,Bar03}. Strong emission in the Balmer
lines and also in other non-hydrogen elements is considered to
be an indicator of youth because these lines are mainly formed 
due to the accretion process and magnetic stellar activity
\citep{Edw94,Muz98}.
As spectroscopic observations are time consuming and sometimes
are restricted to bright objects, a good strategy is to perform
spectroscopic follow-ups only for the most reliable YSO candidates
selected using photometric methods. 
With the public release of all-sky photometric surveys at
different wavelengths, the need of efficient and reliable
techniques to identify and improve the ``automatic'' photometric
classification of YSOs has led to an increasing amount of work
in this field.
The shapes of the spectral energy distributions (SEDs) of young
stars may, in principle, be used for separating YSOs from main
sequence stars. The circumstellar gas and dust (disk and/or envelope)
produces an excess of infrared (IR) emission that can be used
as a diagnostic tool \citep[e.g.][]{All04,Har05}. IR color-color
diagrams have been widely used to distinguish between different
types of YSOs in different star-forming regions \citep[see][as recent
examples]{Gut09,Luh10,Reb11,Koe12,Meg12,Spe13,Stu13,Bro14,Esp14}.
The accretion of this surrounding material onto the surface of
the star produces intense ultraviolet (UV) line and continuum
emission observed also as an excess over the predicted
photospheric emission. The combination of ultraviolet and
infrared photometry has proven to be an useful tool for
identifying YSOs \citep{Fin10,Rod11,Rod13,Gom14}. 
However, an unambiguous identification of YSOs only on
the basis of color-color diagrams is far from being a trivial
task because, among other issues, of contamination from
non-YSO sources such as unresolved background galaxies. A
multi-wavelength fit to the full observed SED using a Bayesian
approach \citep[as in][]{Bar13,Sar14} seems to be a
promising alternative to object-by-object spectroscopic
follow-up. However, in general, SED fitting (and consequently
color-color diagrams) may be highly degenerate between evolutionary
stage and the geometry of the star-disk system \citep{Rob07}.

The Galaxy Evolution Explorer mission \citep[GALEX,][]{Mar05}
has provided images in the far-ultraviolet (FUV, 1350-1780 \AA)
and near-ultraviolet (NUV, 1770-2730 \AA) bands.
The GALEX All Sky Imaging Survey (AIS)
has covered a large part of the sky and all its
products are available online, providing a unique opportunity
to carry out extensive and systematic searches for UV-emitting
YSO candidates \citep{Rod13}. In general, GALEX avoids the
Galactic plane and fields containing bright UV sources
that could damage its detectors and, unfortunately, these
are precisely the fields in which YSOs are most likely to be
found. Despite this, GALEX data are especially useful
to study the UV properties of the dispersed populations of
young stars around star-forming regions \citep[e.g., in
Orion,][]{Bia12} that may be caused by the drift of these
stars away from their formation sites \citep{Fei96}.
In this work, we have undertaken a search for UV-emitting
young stars that are suitable targets to be observed with the
next UV space telescope WSO-UV \citep{Sac14,Shu14}. We are
building GALEX-based catalogs of YSO candidates over large areas
of different star-forming regions \citep[see][]{Gom11,Gom14}.
Here we report our results toward the Orion region.

\section{The GALEX-AIS survey toward Orion}

\begin{figure*}
\centering
\resizebox{\hsize}{!}{\includegraphics{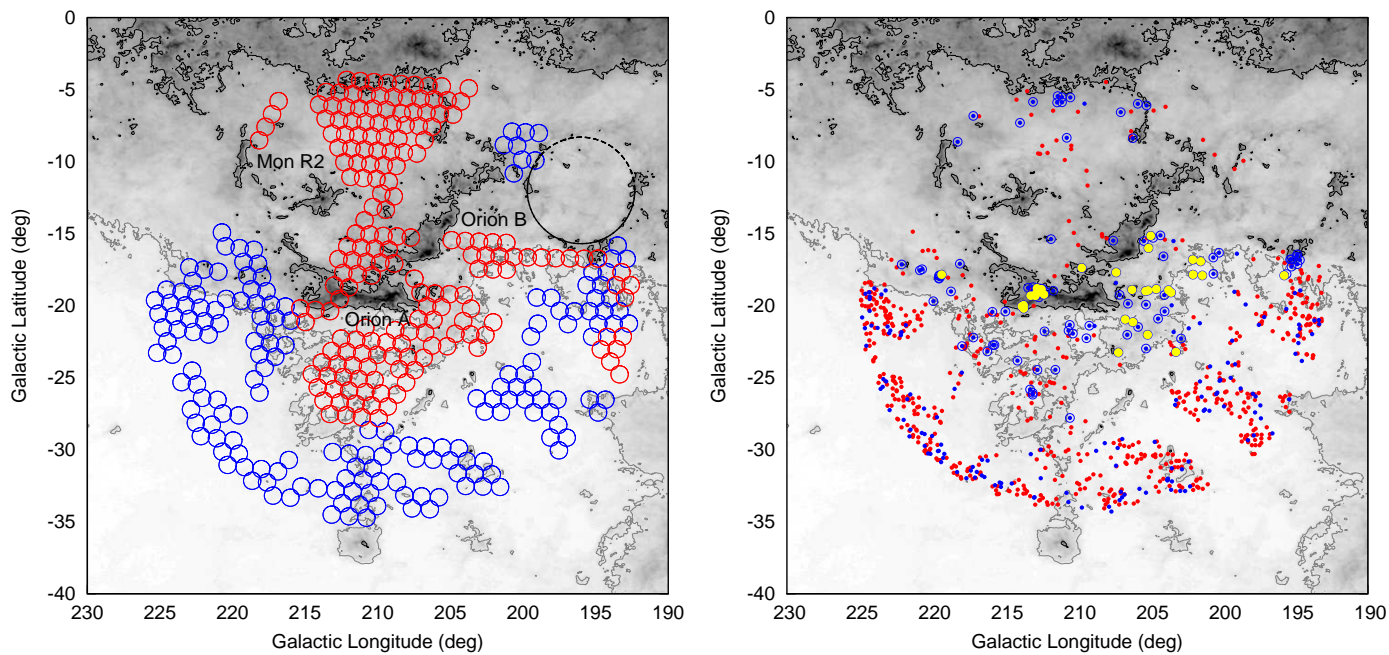}}
\caption{Left panel: GALEX-AIS tiles toward Orion overlaid on
a map taken from the Planck Legacy Archive (HFI 545 GHz data).
The map is in logarithmic gray scale from $\sim 0.1$ MJy/sr
(white) to $\sim 100$ MJy/sr (black) and, for reference, it
has two contours drawn at 2 MJy/sr (gray) and at 10 MJy/sr
(black). Circle radii nearly correspond to the GALEX field
of view (0.6 deg). Blue circles indicate pointings with
data both in FUV and NUV bands, whereas red circles refer
only to the NUV band. The main known star-forming regions
within the studied area (Orion~A, Orion~B and Mon~R2) are
indicated as well as the $\lambda$ Orionis ring (dashed
circle) of clouds.
Right panel: Distribution of sources that satisfy
Equation~\ref{ec_rectangulo} (inside the rectangle
of Figure~\ref{fig_colcol}) overlaid on the same
Planck map. Blue dots are sources that do not fulfill
criteria of likely extragalactic source (any of the
Equations~\ref{ec_koeA}-\ref{ec_koeC}) whereas red
dots are extragalactic contaminants. Blue circles
surrounding the dots represent the final selection
of 111 ``good'' candidates (see text).
There are 30 sources previously identified
as YSOs by others authors (yellow circles) from
which 14 were classified as Classical TTSs.}
\label{fig_tiles}
\end{figure*}

We first retrieved all the GALEX-AIS tiles centered in Orion
with a search radius of $r=15$ deg. This yielded a total of 359
tiles that can be seen in Figure~\ref{fig_tiles} overlaid on a
thermal dust emission map from the Planck mission \citep{Pla11}.
From them, there are 173 tiles with observations both in
FUV and NUV bands whereas 185 tiles have observations in
NUV but not FUV.
Most of the exposure times are in the range
$\sim 50-250$ sec, although some tiles have higher
exposures times but always below $\sim 450$ sec. The
limiting magnitude (at $5 \sigma$ level) for the AIS
Survey with 100 sec is 19.9 mag (FUV band) and 20.8 (NUV)
\citep{Mor07}.
For each tile, we used the band-merged source catalog (data
product ``xd-mcat.fits'') which combines data in one table 
containing all the extracted FUV and NUV sources matched
to the best candidate from the other band. Sources with
signal-to-noise ratio smaller than 2 or separation greater
than 3 arcsec are not matched but are also listed in the
merged catalogs. We joined all these catalogs to make one
single list containing the positions, the calibrated FUV
and NUV magnitudes and their corresponding errors for
1,555,174 GALEX sources in the studied region.

In order to discard spurious sources that may be present in
the GALEX point source catalog \citep{Bia11} we cross-correlated
these sources with the 2MASS catalog \citep{Skr06}. We searched
the 2MASS All-Sky Point Source Catalog within a radius of 15 
degrees around Orion for objects with good photometric
quality\footnote{By ``good'' quality we mean having valid
measurements in the JHK bands, with signal to noise ratio
higher than 5 and magnitude uncertainties smaller than
0.21714, that is having a photometric quality flag of C
or better in the 2MASS catalog. We also discarded sources
with flags indicating possible contamination and/or confusion.}
and we obtained a total of 3,086,388 sources.
The 2MASS point source catalog is complete down to
$J \leq 15.8$, $H \leq 15.1$ and $K \leq 14.3$ mag, although
fainter sources are also included in the catalog. Generally
speaking GALEX-AIS survey would be shallower than the 2MASS
survey for Galactic sources because extinction prevents UV
radiation to propagate large distances in the Galactic plane.
Thus, we expect that no UV-emitting candidates are lost
during the cross-correlation. We consider as ``reliable''
objects those GALEX sources having measurements in FUV
and/or NUV bands and also having a 2MASS counterpart
within a matching radius of 3 arcsec \citep[see][]{Bia11,Gom11}.
The total number of obtained sources is 290,717; that is about
18.7~\% of the initial number of GALEX sources. 
We do not know the true nature of the $\sim80$~\%
of the UV sources that did not match with any 2MASS
source. Point sources listed in the GALEX band-merged
catalog have signal-to-noise ratios as low as $\sim2$,
so that many of these sources may probably be spurious.
However, this seems to be a very large number for spurious
detections when compared with the $\sim 90$~\% of expected
reliability achieved in the AIS at $NUV\sim 22$ according to
GALEX documentation. An empirical test about the reliability
of the GALEX point source catalog is, although interesting,
out of the scope of this paper.

To improve the photometric information we have also
cross-matched the sources with other publicly
available catalogs with, again, a matching radius of 3
arcsec. We have found significant matches (i.e. point
sources unaffected by known artifacts) in the UCAC4
\citep{Zac13}, SDSS \citep{SDSS}, DENIS \citep{DENIS},
CMC15 \citep{CMC15}, and WISE \citep{Wri10} catalogs.
Thus, apart from GALEX ($FUV$,$NUV$) and 2MASS ($JHK$) photometric
data we added, when available, magnitudes and errors in $BgVri$
from UCAC4, $ugriz$ from SDSS, $I$ from DENIS, $r$ from CMC15
and $W1-4$ from WISE.

\section{Young star candidate selection}

We are particularly interested in TTSs, so we performed a
search for this type of objects in the 290,717 sources
of the sample. We used the SIMBAD database to search for
confirmed (not candidate) TTSs and we found 56 stars.
We have addressed the candidate selection process from
an empirical point of view, by preparing different combinations
of color-color diagrams and looking at the distribution of the
sources in each diagram. After many different tests
we concluded that the $NUV-J$
versus $J-W1$ diagram gives reasonably good results, at
least for the sample we are working with. On the one hand,
$NUV-J$ is a color index that may indicate UV excess and that
has been previously used by other authors for searching and
finding active low-mass stars \citep[e.g.,][]{Fin10,Rod11,Shk11}.
On the other hand, the IR excess produced by the presence of
protoplanetary disks in Classical TTSs can be quantified
by $J-W1$. The utilization of different color indices such
as $J-K$ \citep[as in][]{Fin10,Rod11} or $J-W2$ \citep{Rod13}
shifted stars with infrared excess closer to or further away
from the main sequence but did not result in any significant
difference in the selection procedure.

\begin{figure}
\centering
\resizebox{\hsize}{!}{\includegraphics{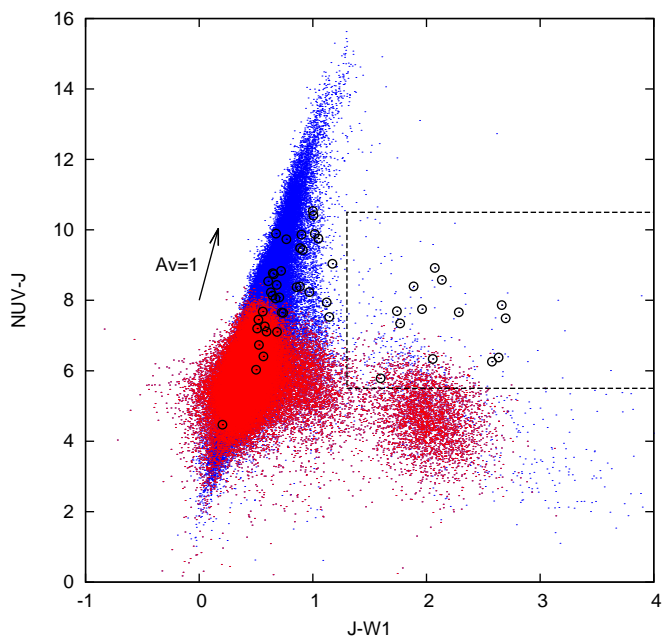}}
\caption{$NUV-J$ versus $J-W1$ for stars in the sample.
Blue dots refer to the full sample of stars toward Orion
and red dots refer to extragalactic sources according
to the used criteria (see text). Open circles are
identified TTSs in the SIMBAD database, The $A_V=1$
mag reddening vector is shown with an arrow. Dashed
line indicates the region used for a first selection
of TTS candidates.}
\label{fig_colcol}
\end{figure}

In Figure~\ref{fig_colcol}, the $NUV-J$ vs. $J-W1$ diagram for
the 290,717 sources of the sample is shown.
Most of the SIMBAD-identified sources (all of them except
one) are located in the range $5.5 \la NUV-J \la 10.0$,
whereas the infrared excess exhibits two well-defined
groups separated at $J-W1 \simeq 1.3$. We have made a
first list of candidates by selecting the region of the
reddest known TTSs (dashed line in Figure~\ref{fig_colcol}),
that is by using the following criteria:
\begin{equation}
\label{ec_rectangulo}
\begin{array}{l}
5.5 \leq NUV-J \leq 10.5 \\
1.3 \leq J-W1 \ .
\end{array}
\end{equation}
There is a group of less reddened TTSs ($J-W1 \leq 1$),
probably including Weak-lined TTSs, that can not be
separated from main sequence objects in this
color-color diagram. To extend the selection criterion
to smaller $J-W1$ values would include these TTSs but
also a lot of main sequence stars and other contaminants
that are difficult to separate at a later stage. We have
decided to keep a more restrictive criterion
(Equations~\ref{ec_rectangulo}) at this stage to ensure
a more reliable list of TTS candidates and mitigate
contamination. Despite this, it is necessary to remove
contamination by non-YSO sources by applying additional
criteria. For this we adopt the photometric scheme based
on WISE colors developed by \citet{Koe12}. They classified
as unresolved star-forming galaxies those satisfying all
the following constraints:
\begin{equation}
\label{ec_koeA}
\begin{array}{rcl}
W1-W2 & < & +0.46 \times \left( W2-W3-1.7  \right) \\
W1-W2 & > & -0.06 \times \left( W2-W3-4.67 \right) \\
W1-W2 & < & -1.00 \times \left( W2-W3-5.1  \right) \\
W1-W2 & > & +0.48 \times \left( W2-W3-4.1  \right) \\
   W2 & > & 12 \\
W2-W3 & > & 2.3 \ ,
\end{array}
\end{equation}
whereas unresolved AGNs are those sources for which either
\begin{equation}
\label{ec_koeB}
\begin{array}{rcl}
W2 & > & +1.9 \times \left( W2-W3+3.16  \right) \\
W2 & > & -1.4 \times \left( W2-W3-11.93 \right) \\
W2 & > & 13.5
\end{array}
\end{equation}
or
\begin{equation}
\label{ec_koeC}
\begin{array}{rcl}
W1 & > & +1.9 \times \left( W1-W3-2.55 \right) \\
W1 & > & 14.0 \ .
\end{array}
\end{equation}

\begin{figure}
\centering
\resizebox{\hsize}{!}{\includegraphics{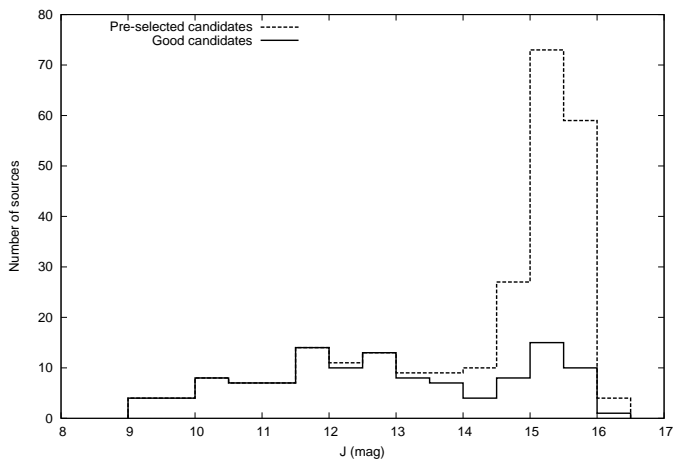}}
\caption{Magnitude histogram in the 2MASS J band
for all the pre-selected TTS candidates (dashed
line) and for the selection of ``good'' candidates
(solid line) as explained in the text.}
\label{fig_histoJ}
\end{figure}

There are 83,484 objects fulfilling the extragalactic
contamination criteria (Equations~\ref{ec_koeA} or
\ref{ec_koeB} or \ref{ec_koeC}) that are indicated
as red dots in Figure~\ref{fig_colcol} and were
excluded from the TTS candidate selection criteria
(Equations~\ref{ec_rectangulo}). The remaining 259 sources
are shown in the right panel of Figure~\ref{fig_tiles} as
blue dots, whereas red dots in this figure represent
unresolved galaxies and AGNs according to the criteria
of \citet{Koe12}. We note that the latter are distributed
preferentially along the lower semicircumference of the 
studied region where there is no or very little dust
present. The pre-selected candidates (blue dots) that
are located in this region are precisely the faintest
objects, with apparent magnitudes in the range
$14.5 \la J \la 16$ (as shown in Figure~\ref{fig_histoJ}).
Thus, it is likely that at least part of these sources are
background extragalactic contaminants that are still present
in the sample because, obviously, these color/magnitude
selection criteria are not perfect and there can be source
misclassification either including some extragalactic objects
or removing some YSOs \citep[see Figures~7 and 8 in][]{Koe12}.
Instead of making a cut in magnitude we use the spatial
distribution to perform a final cleaning of the sample
keeping the faintest TTS candidates. Thus, from the 259
sources fulfilling the above mentioned criteria (blue dots
in right panel of Figure~\ref{fig_tiles}), we select as ``good''
TTS candidates (blue circles surrounding the dots in
Figure~\ref{fig_tiles}) those sources that are not
located along or close to the semicircumference
with a relatively high density of extragalactic sources.
We also excluded nine objects that Sloan classifies
as extended sources and the object 2MASS J06235520+0018433
because it is a quasar discovered by \citet{Im07}.

Our final list of 111 TTS candidates is shown in
Table~\ref{tab_TTSsample}. There are 30 of these objects
that were previously classified as YSOs or YSO candidates by
other authors; in these cases the specified object type
and the corresponding reference are included in last columns
of Table~\ref{tab_TTSsample}. Of these 30 previously identified
objects, 14 have been classified as Classical TTS whereas 13,
although probable Classical TTS, have been generically classified
as TTS or YSO. There are only 3 sources that were classified as
Herbig Ae/Be objects.
The remaining 81 sources are new TTS candidates identified
in this work. We expect that most of these candidates are
Classical TTS (they have UV and IR excesses).
\begin{table*}
\caption{List of the 111 TTS candidates.}
\label{tab_TTSsample}
\centering
\begin{tabular}{cccccccc}
\hline
\hline
ID(2MASS) & $l$ (deg) & $b$ (deg) & $NUV$(GALEX) & $B$(UCAC4) & $J$(2MASS) & $W1$(WISE) & Ref (known) \\
(col 1) & (col 4) & (col 5) & (col 8$\pm$9) & (col 12$\pm$13) & (col 34$\pm$35) & (col 40$\pm$41) & (col 49) \\
\hline
 J05071385-1020045 & 210.512 & -27.796 & 21.097$\pm$0.416 & (...)           & 15.306$\pm$0.058 & 13.671$\pm$0.027 &       \\
 J05113654-0222484 & 203.205 & -23.212 & 17.035$\pm$0.019 & 12.698$\pm$0.17 & 10.558$\pm$0.022 &  7.87 $\pm$0.023 & Lee07 \\
 J05141589-0138583 & 202.856 & -22.279 & 19.251$\pm$0.098 & 13.58 $\pm$0.04 & 10.108$\pm$0.024 &  8.752$\pm$0.023 &       \\
 J05160265-0356398 & 205.273 & -22.976 & 21.751$\pm$0.854 & (...)           & 15.362$\pm$0.065 & 13.875$\pm$0.028 &       \\
 J05160402+0618525 & 195.751 & -17.904 & 18.787$\pm$0.074 & 15.532$\pm$0.33 & 11.286$\pm$0.022 &  8.683$\pm$0.023 & Tak10 \\
\hline
\end{tabular}
\tablefoot{This table is published in its entirety in the electronic edition.
A portion is shown here for guidance regarding its form and content.}
\end{table*}


\section{Properties of the selected TTS candidates}

\subsection{Spatial distribution}

A detailed analysis of the spatial distribution is difficult
because of the bias in the non-uniform GALEX pointing distribution
toward Orion. For instance, the lack of sources in the range
$-15 \la b \la -10$ deg is likely more related to the lack of
pointings than anything else (see Figure~\ref{fig_tiles}).
Despite this, we note that in general sources are spread almost
uniformly except for two apparent overdensities. The first one
is located at galactic coordinates $l \simeq 195$ deg and
$b \simeq -17$ deg corresponding to one of the clouds of
the  $\lambda$ Orionis ring of clouds (B~223). The second
overdensity is at $l \simeq 213$ and $b \simeq -19$ deg
toward the Orion A cloud and it agrees with the cluster
rich in Class III stars found by \citet{Pil13} in the
southern end of L1641. These apparent overdensities
are seen in regions in which GALEX-AIS covered a
dense part of these molecular clouds (B~223 and Orion~A,
see left panel in Figure~\ref{fig_tiles}). At this point 
we can only say that there seems to be a relatively high 
density of sources toward denser and dust-richer regions 
embedded within an almost more or less homogeneous 
distribution of TTS candidates. This picture is consistent 
with the result of \citet{Bia12} concerning the existence 
of a mixture of a young clustered population and a 
widespread population originated from an earlier episode
of star formation. However, this scenario is a little
uncertain because of the presence of a rich foreground
population of young ($\sim 5-10$ Myr) stars toward
Orion \citep{Bou14}.

\subsection{Spectral energy distributions}

\begin{figure}
\centering
\resizebox{\hsize}{!}{\includegraphics{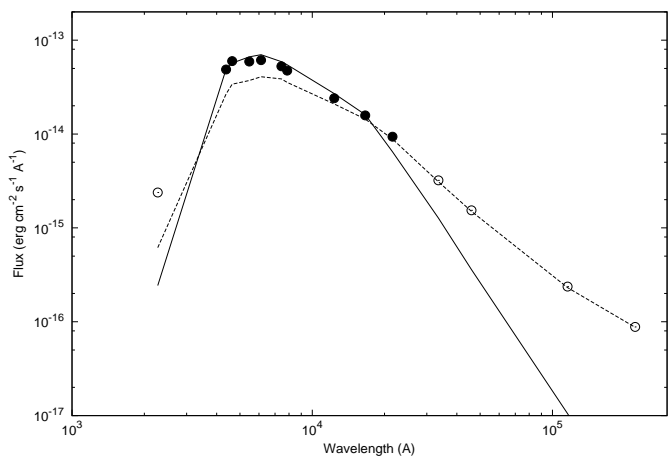}}
\caption{Observed and best fitted flux densities for one
example source, J05191356-0324126, for which we
obtained $T_\mathrm{eff}=4900$~K, $A_V = 0.5$ and that it is
a Classical TTS according to \citet{Lee07}. Dashed line
indicates the observed photometric data. Circles are
dereddened data points: solid circles are considered
in the fitting process whereas open circles show
ultraviolet and/or infrared excesses. Solid lines
indicate the best BT-Settl fitted model. Error
bars for this source are smaller than point sizes.}
\label{fig_seds}
\end{figure}

The SEDs of the TTS candidates were analyzed by using the
VOSA web-tool developed by the Spanish Virtual Observatory
\citep{Bay08}. VOSA performs a $\chi^2$ statistical
test to determine the theoretical spectral model that
best reproduces the observed data. VOSA has a flexible
environment allowing the user to choose among different
available collections of models and to define the
parameters (and their ranges) to be fitted \citep[see
details in][]{Bay08}. 
We have used the available photometry (magnitudes
and errors in Table~\ref{tab_TTSsample}) for fitting
the observed SEDs allowing veiling (UV excess)
at wavelengths shorter than 3000 \AA\ (i.e.
in the GALEX FUV and NUV bands) and infrared excess
at any of the WISE bands.
The distance to the sources is a fixed value
in VOSA and mandatory to calculate the bolometric
luminosities. The distance to the well-studied
Orion Nebular Cluster (ONC) seems to be $\sim 400$
pc \citep{San07,Men07}. However, when dealing with
a very sparse population distance variations may be
significant. \citet{Lom11} obtained a distance of
$\sim 370$ pc for Orion A and $\sim 400$ pc for
Orion B. \citet{Sch14} obtained larger values and
significant distance variations ($490 \pm 50$ pc)
toward different lines of sight suggesting a complex
3D structure in Orion. In fact, it seems that the
eastern edge of Orion is $\sim 70$ pc farther from
us than the ONC. Since VOSA allows us to provide
a value for the distance error, which is propagated
to the derived luminosity, we adopted a distance
of $400 \pm 20$ pc to all the sources. The SED
fitting process is quite insensitive to some
parameters such as, for instance, the surface 
gravity of the source \citep{Bay08}. Thus, in
order to improve the convergence of the fitting
procedure we fixed both metallicity and surface
gravity around the expected values (solar
metallicty and $\log g = 4$). We leave the
visual extinction as a free parameter constrained
to the range $0.0 \leq A_V \leq 5.0$ and the
effective temperature of the theoretical model
is left completely free.
There were 65 sources that converged to physically
reasonable solutions and their SEDs were well-fitted
by the BT-Settl models \citep{All12}.
Figure~\ref{fig_seds} shows the result
for an example fitted SED.
Most, though not all, the fitted SEDs displayed
significant UV and IR excesses and we consider these
sources as very good candidates to TTSs.

\subsection{Derived physical properties}

The effective temperature ($T_\mathrm{eff}$), bolometric luminosity
($L_{bol}$) and visual extinction ($A_V$) derived for each
object from their fits are shown in Table~\ref{tab_FITsample}.
\begin{table*}
\caption{Physical properties derived from the best SED fits 
{\bf for 65 sources.}}
\label{tab_FITsample}
\centering
\begin{tabular}{ccccccccc}
\hline\hline 
 ID(2MASS) & $l$ & $b$ & $T_\mathrm{eff}$\tablefootmark{a} & 
           $L_\mathrm{bol}$ & $\Delta L_\mathrm{bol}$ & $A_V$ & Type & Ref. \\
           & (deg) & (deg) & (K) & ($L_{\odot}$) & ($L_{\odot}$) & (mag) & \\
\hline
 J05071385-1020045 & 210.512 & -27.796 & 4700 & 0.035 & 0.003 & 1.0 \\
 J05141589-0138583 & 202.856 & -22.279 & 4400 & 2.355 & 0.275 & 0.0 \\
 J05160402+0618525 & 195.751 & -17.904 & 3600 & 1.343 & 0.155 & 1.25&TTS&Tak10\\
 J05170743-1143036 & 213.069 & -26.18  & 4200 & 0.029 & 0.003 & 0.0 \\
 J05173566-1140050 & 213.073 & -26.054 & 3800 & 0.024 & 0.003 & 0.0 \\
\hline
\end{tabular}
\tablefoot{This table is published in its entirety in the electronic edition.
A portion is shown here for guidance regarding its form and content.\\
\tablefoottext{a}{Uncertainty in $T_\mathrm{eff}$ is $\pm$50~K.}}
\end{table*}

Last two columns give the object type and
reference for previously identified stars.
Given the large uncertainties in their distances
the derived luminosities must be taken with caution.
However, the calculated temperatures allow us to
confirm the cool photospheres expected for TTSs.
Most of the sources (58) have stellar effective
temperatures $T_\mathrm{eff} \la 5000$~K (spectral type
K0 or later). There is only one discrepant case
(J05401176-0942110) for which we obtain
$T_\mathrm{eff} \simeq 2600$~K but that, according
to \citet{Man06}, is a probable Herbig Ae/Be star.
We do not notice any particular behavior in the derived
physical properties or in their spatial distribution,
although obviously there are too few data points to draw
any robust conclusion in this sense.

\section{Conclusions} 

We constructed a catalog of 111 reliable TTS candidates
with detected UV emission in an area of $\sim$ 400 square
degrees toward the Orion star-forming region, from which
81 sources are new candidates identified in this work.
We derived physical properties for 65 of these candidates.
Most of the sources show photospheric temperatures and
both UV and IR excesses compatible with the expected for TTSs.
These TTS candidates likely belong to a dispersed population
of young stars, as also observed on other star forming
regions \citep[e.g.][]{Com13}, whose origin remains unclear.
Their study will allow further understanding of the physical
properties and origin of this population of young stars.

\begin{acknowledgements}
We thank the referee for his/her comments which improved this paper.
We acknowledge financial support from Ministerio de Econom\'ia
y Competitividad of Spain through grant AYA2011-29754-C03-01.
\end{acknowledgements}

\end{document}